\documentclass[preprint,prb,showkeys,showpacs,tightenlines]{revtex4}
\usepackage{graphicx}
\begin{document}
\preprint{Bicocca-FT-03-35;  December 2003}

\title
 {Updated tests of scaling and universality for the spin-spin \\
  correlations in the 2D and 3D spin-$S$ Ising models\\
   using high-temperature expansions\\}

\author{P. Butera\cite{pb} and M. Comi}
\affiliation
{Istituto Nazionale di Fisica Nucleare and\\
Dipartimento di Fisica, Universit\`a di Milano-Bicocca\\
 3 Piazza della Scienza, 20126 Milano, Italy}
\date{\today}
\begin{abstract}

We have extended, from order 12 through order 25, the high-temperature series 
expansions (in zero magnetic field) for the spin-spin correlations 
of the  spin-$S$ Ising models 
 on the square, simple-cubic and body-centered-cubic lattices.
On the basis of this large set of data, we confirm accurately      
the validity of the scaling and universality hypotheses 
 by resuming several tests which involve
the correlation function, 
 its moments and  the exponential or the second-moment correlation-lengths. 
\end{abstract}

\pacs{ PACS numbers: 05.50+q, 11.15.Ha, 64.60.Cn, 75.10.Hk}
\keywords{critical behavior,Ising model, scaling, universality, universal 
combinations of critical amplitudes, high-temperature expansions}

\maketitle

\section{Introduction and conclusions}

Moderate-length high-temperature (HT) expansions (through order 12) 
 and low-temperature (LT) expansions 
for the  spin-spin correlation function (sscf) $G( \vec r,T;S)$ 
of the nearest-neighbor  
Ising models with general spin S were first 
computed\cite{camp,camp2,foxgu} three decades ago 
on various lattices in 2D and in 3D. 
Motivations  for the study  of these models came not only from 
their direct phenomenological interest, but mainly
 from the conjecture\cite{dosy} that, in a given space dimension,  
the exponents characterizing the
 critical behavior are independent both of the lattice structure and 
 of the spin magnitude $S$. This conjecture was 
the first step towards the modern notion of  universality class.   
In the same years also the 
hypothesis of critical scaling\cite{widom} was put forward.
 Many  
studies\cite{camp,camp2,foxgu,fishbur,fmw,mjw,fw,fermor,tarfish,fishtar,fer} 
of the mentioned HT and LT series 
were devoted to 
test the validity and the  main consequences of these  
basic hypotheses\cite{univ,widom,fishbur,fermor,dg,wil,pok,dc,mefrmp,dosy}.  
 Although  the results  sometimes were not as precise as was hoped,  
or covered only the   $S=1/2$ case, 
 the scaling tests suggested  that the critical sscf is a homogeneous
 function of appropriate variables, while the 
universality tests indicated that 
 the critical indices 
 and suitable combinations\cite{aha} of critical amplitudes
 are independent of the spin $S$ and lattice structure.
 A few years later, the first substantial extension\cite{nick21,nr90} 
of   HT Ising series in 3D
(through order 21 on the body-centered-cubic(bcc) lattice only) 
  did not make higher  expansion coefficients available for the sscf, 
but only for its two lowest even moments
 and therefore various tests could not be repeated and updated.

We are now resuming the HT part of those pioneering analyses  
in order to improve their extent and accuracy by taking advantage 
of our recent extension\cite{bcesse25,bcol,bcesse2d25} from order $12$ 
through order $25$ of the HT expansions for the sscf  of the 
Ising model  with general spin $S$,
 in 2D on the square (sq) lattice  and in 3D on the 
simple-cubic (sc)  and the bcc  lattices.
 From these data we have also derived series for related 
  quantities, in particular for a variety of moments of the sscf, 
which are computed through order 25, and 
 for the exponential (or `true') correlation
 length defined via the exponential decay of the sscf, 
which, however, can be extended only through order 19.
 For reasons of space we have not tabulated  in this paper 
the  series analysed, but  have  included them  
into our on-line library\cite{bcol}  
of HT data for the spin-$S$ Ising model in order to make  them more widely 
available for further study.
Since this is the largest body of series data so far computed 
for these systems,
 we have already been studying other aspects of them in previous papers. 
In particular, in Ref.\cite{bcesse25} we have accurately confirmed
 that the residual  weak  spin-dependence 
observed\cite{zinn81} in lower order studies 
 of the susceptibility   exponent $\gamma$
and of the correlation-length exponent   $\nu$ in 3D on the bcc lattice,
should not be ascribed to  small violations of
universality, but can be simply explained away as  numerical inaccuracies 
 due to  expected non-negligible 
 spin-dependent corrections to the leading scale behavior.
Moreover we have tested  the universality 
of several amplitude combinations obtaining 
 similar results. In Ref.\cite{bcesse2d25} an
 analogous  survey
 of  universal quantities was performed in 2D for the sq lattice case. 
 Shorter series  (but only for the $S=1/2$ case)
 had been analysed in Refs.\cite{older}. 

From the evidence presented here we can conclude  that   our HT data
 for the sscf have by now 
reached an extension sufficient to  make the use of modern
 series-extrapolation techniques possible and generally reliable.
 Therefore we are able to   exhibit more convincingly
  both in 2D and in 3D 
 many expected  properties related to scaling and universality also in  
 some cases in which the old analyses 
 led to   inconclusive or not very precise  results.

 The rest of the paper is organized as follows. In section II
we shall outline  the main features of the model, 
introduce our notations and conventions
 and very briefly recall the scaling and universality properties expected for 
 the sscf along with the corresponding tests discussed in full detail 
by the above cited 
 papers\cite{fishbur,fmw,mjw,fw,tarfish,fishtar,fer}.
Therefore, in section III we can restrict ourselves 
to only a few comments on the numerical results.

\section{ The spin-$S$ Ising models}

  The spin-$S$ Ising models  with nearest-neighbor 
interaction are defined by  the Hamiltonian:
\begin{equation}
H \{ s \} = -{\frac {J} {2}} \sum_{( \vec r,{\vec r}') } 
s({\vec r})  s({\vec r}') -h\sum_{\vec r} s({\vec r})
\label{hamilt} \end{equation}
where $J$ is the exchange coupling, and
  $s(\vec r)=s^z(\vec r)/S$  with  $s^z(\vec r)$  a  
classical spin  variable at the
lattice site $\vec r$, taking the $2S+1$ values 
$-S, -S+1, \ldots,S-1, S$.    
 The sum runs over  all nearest-neighbor pairs of  sites.
 For simplicity, the nearest-neighbor lattice spacing will be set equal
 to $1$ everywhere. 
We shall 
 consider expansions  
in the usual HT variable $K=J/k_BT$   
   where $T$ is the temperature, $k_B$
 the Boltzmann constant, and $K$ will be called
``inverse temperature'' for brevity. 
In the critical region we shall also refer to 
 the standard `reduced-temperature' variable 
 $t(S)=1-T_c(S)/T = 1 - K/K_c(S)$.

We shall study the HT expansion of the (connected) sscf defined as 
\begin{equation}
G( \vec r,T;S)= \langle s(\vec 0)  s(\vec r) \rangle_c
\label{corf} \end{equation}
In order to estimate numerically $G( \vec r,T;S)$ as $T \rightarrow T_c+$ , 
we have  allowed for  its expected\cite{fishbur,onsa} 
behavior: in the 2D case
\begin{equation}
 G( \vec r,T;S) \approx   G(\vec r,T_c;S)- E^+(\vec r;S) t(S) 
{\rm ln} t(S) +\ldots
\label{singpart2d}\end{equation}
and in the 3D case
\begin{equation}
G( \vec r,T;S) \approx G(\vec r,T_c;S)- E^+(\vec r;S) t(S)^{1-\alpha} +\ldots
\label{singpart3d}\end{equation}
Here $E^+(\vec r;S)$ is the critical amplitude 
of the leading singular correction,
 $\alpha=0.110(1)$\cite{bcesse25} denotes the critical exponent 
of the specific heat in 3D and the dots
indicate higher order corrections.

The correlation-function moment $ \mu_{n}(T;S)$ of order $n$
   is defined as
\begin{equation}
 \mu_{n}(T;S)=\sum_{\vec r} |\vec r|^n \langle s(\vec 0)  s(\vec r) 
\rangle_c 
\end{equation}
(for $n<0$ the sum extends to $\vec r \not= 0$).

The expected asymptotic behavior of $\mu_{n}(T;S)$ as $T \rightarrow T_c+$ is
\begin{equation}
 \mu_{n}(T;S) 
\approx m_n^+(S)t(S)^{-(\gamma+n\nu)}[1 + a_{n}^+(S) t(S)^{\theta}+ \ldots].
\label{asbsm} \end{equation}

In 2D the exponent $\theta$ of the leading singular correction
 is larger than unity, while 
 in 3D a recent simultaneous study\cite{blode} of a set of models 
in the Ising universality
class has suggested the very precise estimate $\theta= 0.517(4)$.

The scattering function, namely the Fourier transform of $G(\vec r,T;S)$
\begin{equation}
\hat G(\vec k, T;S) = \sum_{\vec r} exp[-i\vec k \cdot \vec r] G(\vec r, T;S)
\end{equation}
 for $\vec k=0$ yields  the zero-field reduced susceptibility 
\begin{equation}
\hat G(\vec 0, T;S) = \mu_{0}(T;S) = \chi(T;S) = \sum_{\vec r} 
\langle s(\vec 0)  s(\vec r) \rangle_c  
\label{chi} \end{equation}
The second-moment correlation-length is defined in $d$ spatial dimensions by 
\begin{equation}
 \xi^{2}_{sm}(T;S)=  
 \frac  {\mu_{2}(T;S)} {2d\chi(T;S)}=
\frac {d {\rm ln} \hat G(\vec k, T;S)}{dk^2 }|_{k=0} 
\label{smcl} \end{equation}	
 For $T>T_c$ the sscf is exponentially decreasing for large
$r$ and therefore
following Ref.\cite{fishbur}, 
 beside the `second-moment' correlation-length 
we can also define the  inverse `exponential' (or `true')
correlation-length in the direction $\vec e$ as  

\begin{equation}
k_{\vec e}(T;S)= 
-lim_{r \to \infty} \quad \frac{1} {r} {\rm ln} | G(r \vec e,T;S) |
\label{tcl} \end{equation}

Since  the singularity of $\hat G(\vec k, T;S)$
 closest to the real axis in the complex $ k$ plane  is located at 
 $\pm i k_{\vec e}(T;S)$, the exponential correlation-length
 can  be obtained by solving recursively\cite{fw}
 the eq.
\begin{equation} 
\hat G(i\vec k_{\vec e}, T;S)^{-1}=0.
\end{equation}
Rather than working  directly with $k_{\vec e}(T;S)$ 
 which is not an ordinary power series in $K$, 
it is  expedient\cite{fishbur} to form the quantity 
\begin{equation}
\xi^2_{\vec e}(T;S) = \frac{f^2} { 2 [{\rm cosh} (fk_{\vec e}) - 1]}
\label{mtcl}\end{equation}
which is an ordinary power series in $K$. In Eq.(\ref{mtcl})
 $f$ is a geometrical factor depending on the unit vector  ${\vec e}$
 and on  the 
lattice considered.  In particular, if $\vec e$ is directed along
 a lattice axis, we have $f=1$ for the sq and the sc lattices, while  
$f=1/\sqrt 3$ for the bcc lattice.
 
 So far,  3D  data for this quantity were published   
exclusively  for $S=1/2$, and did not extend  beyond order 15
in the sc lattice case\cite{pisa} or
beyond order 10 in the bcc lattice case\cite{fw}.
In 2D the HT expansion can be computed exactly \cite{fishbur,onsa} for 
$S=1/2$, but no data have been published for $S \neq 1/2$.
In Ref.\cite{bcol}, we have  tabulated the 
expansion of $ \xi^2_{\vec e}(T;S)$ through
 order 19  for ${\vec e}$ directed along a lattice axis, in the case
 of the sq, sc and bcc lattices and with $S=1/2,1,3/2,2,5/2,3,\infty$.

 In order to avoid possible confusion, it should be pointed out
 that in Ref.\cite{fishbur} our $ \xi^2_{\vec e}$   was denoted 
by $\Lambda'_2(\vec e)$, while the symbol $ \xi_{\vec e}$ 
 was used to denote  $k^{-1}_{\vec e}$.
 Our notation might be more suggestive since our 
 $ \xi^2_{\vec e}$ compares very closely with $\xi^2_{sm}$. 
 Indeed, the true and second-moment correlation-lengths 
 are almost identical in magnitude above the critical temperature. 
In particular  on the sq lattice, when $\vec e$ is directed 
along a lattice axis,
 the HT expansion coefficients of  
$\xi^2_{\vec e}$ and $\xi^2_{sm}$ coincide through sixth order
  for $S=1/2$ , through fourth order  
 for $S=1$ and through  second order  for higher values 
of the spin.
In 3D, in the sc lattice case, the expansion coefficients of  
$\xi^2_{\vec e}$ and $\xi^2_{sm}$ coincide through
 seventh order    for $S=1/2$, through fifth  order  for
$S=1$ and through third order   for higher values of $S$.
In the case of the bcc lattice, the expansion coefficients coincide
 through third  order  for all values of the spin. 
Moreover, up to the maximum order of our computation, the noncoinciding 
  coefficients  differ by less than $0.1\%$.


 The two correlation lengths $\xi_{\vec e}$ and $\xi_{sm}$ are 
expected to share the same critical exponent $\nu$
 so that their asymptotic behavior when $T \rightarrow T_c+$ can be 
written as
\begin{equation}
\xi_{sm} (T;S) 
\approx f^+_{sm}(S)t(S)^{-\nu}[1 + a_{sm}^+(S) t(S)^{\theta}+ \ldots]
\label{asbxsm} \end{equation}
 and
\begin{equation}
\xi_{\vec e} (T;S) 
\approx f^+(S)t(S)^{-\nu}[1 + a^+(S) t(S)^{\theta}+ \ldots]
\label{asbxtrue} \end{equation}

Here the critical amplitude $f^+(S)$ is independent of $\vec e$, since the
 sscf becomes spherically symmetric near the critical point. The ratio 
\begin{equation}
Q^+_{\xi}(S)=f^+(S)/f^+_{sm}(S) 
\label{ratio} \end{equation} 
 is a universal combination of  critical amplitudes\cite{aha} 
 i.e. it is  expected to depend only on the lattice dimensionality $d$, 
but not on the spin $S$ or the 
lattice structure.

For $T \rightarrow T_c+0 $ and in zero magnetic field, the sscf
 is expected 
 to exhibit the asymptotic structure
\begin{equation}
G(\vec r,T;S) \approx (1/r)^{d-2+\eta}A_l(S){\rm D_0}\Big ( C_l(S)r/
\xi_{sm}(T;S)\Big ) 
+ \ldots 
\label{scalfun}\end{equation}
when both $r$ and $\xi_{sm}$ are much larger than the lattice spacing 
(with arbitrary $r/\xi_{sm}$). 
 Eq.(\ref{scalfun}) together with these
assumptions on $r$ and $\xi_{sm}$, is usually
 referred to as the ``strong-scaling hypothesis'' 
(while it is called the ``weak-scaling hypothesis'', if 
its validity is restricted to the $r \rightarrow \infty$ limit
with fixed $r/\xi_{sm}$). 
In Eq.(\ref{scalfun}), $\eta$ is the critical exponent 
 describing the decay of the sscf at the critical point,
 ${\rm D_0}(x)$ is called the critical scaling function,
   $A_l(S)$ and  $C_l(S) $ are scale factors.
The dots indicate subcritical corrections 
proportional to a positive power
 of some irrelevant field. 
The scaling function ${\rm D_0}(x)$ is expected to be 
universal: 
its structure does not depend on the particular model under study 
provided that it belongs to a given universality class. 
On the contrary 
 the scale factors $A_l(S), C_l(S) $ depend on the spin and the lattice $l$. 
The validity of the asymptotic structure Eq.(\ref{scalfun})
 was verified analytically\cite{kada} for the
 spin $S=1/2$ Ising model in 2D. 

For the scattering
 function $\hat G(\vec k, T;S)$
 the analogous scaling form as $T \rightarrow T_c+$  can be written as
\begin{equation} 
\hat G(\vec k, T;S) \approx A'_l(S)t(S)^{-\gamma}\hat
 {\rm D}'_0 \Big (C'_l(S)k^2 \xi^2_{sm}\Big )+ \ldots 
\label{gk} \end{equation}
 If the  scale factors $A'_l(S), C'_l(S) $ are 
 specified adopting the normalization conditions
 \begin{equation}
 {\hat {\rm D}'_0}(0)=1; 
  {\quad   \quad \big(\frac {d \hat {\rm D}'_0(x)}{dx}}\big )_{x=0}=-1. 
 \end{equation}
 one can write\cite{fishbur} as $k \rightarrow 0$  
\begin{eqnarray}
 \hat G(\vec 0, T;S)/\hat G(\vec k, T;S)= 1/g_+(k \xi_{sm}(T;S))=\\
1 +  \xi^{2}_{sm}(T;S) k^2  -\Sigma_4(T,S) \xi^{4}_{sm}(T;S) k^4+ 
\Sigma_6(T,S) \xi^{6}_{sm}(T;S) k^6 + O(k^8 )
\nonumber
\label{gkl} \end{eqnarray}
where the function $g_+(k \xi_{sm}(T;S))$ is universal and thus the quantities 
\begin{equation}
\Sigma_4(T,S)= c_4 \frac{\mu_4(T;S)\mu_0(T;S)}{\mu^2_2(T;S)} -1
\label{sigma4} \end{equation}

\begin{equation}
\Sigma_6(T,S)= c_6 \frac{\mu_6(T;S)\mu^2_0(T;S)}{\mu^3_2(T;S)} -
2 \Sigma_4(T,S) -1
\label{sigma6} \end{equation}
with $c_4=1/4$ and $c_6=1/36$ for $d=2$, while $c_4=3/10$ and $c_6=3/70$ for
 $d=3$, 
have finite universal values as $T \rightarrow T_c+$.

All $\Sigma_{2n}(T,S)$,   
as well as  the difference $\xi^2_{\vec e}-\xi^2_{sm}$, vanish\cite{fishbur}
 in the mean-field  related approximations.
Therefore  the magnitudes  of these quantities at 
the critical point 
 can be  taken as  a measure of the deviation 
of a given system from gaussian behavior, which turns out to be very
 small on the HT side of the critical point.
 
More generally, it was observed\cite{fermor} that the scaling hypothesis 
Eq.(\ref{scalfun}) implies that, at the critical point,  the ratios 
\begin{equation}
R_{m,n;r,s}(T;S)=\frac{\mu_m(T;S) \mu_n(T;S)} {\mu_r(T;S) \mu_s(T;S)}
\label{Rmnrs} \end{equation}
with $m+n=r+s$, are universal. These ratios are 
dominated by the critical singularity
also for negative values of the indices  $m,n,r,s$  
 provided that each index exceeds $-2+\eta$, 
as follows from Eq.(\ref{asbsm}).

Finally, the determination of the amplitude $E^+(\vec r;S)$  
of the leading singularity of the sscf (see Eqs.(\ref{singpart2d}) 
and (\ref{singpart3d}))   gives another opportunity to perform 
universality and scaling tests.
In order that  the structure of Eqs.(\ref{singpart2d}) 
and (\ref{singpart3d})  be compatible\cite{fishtar}
 with the strong-scaling hypothesis Eq.(\ref{scalfun}), the amplitudes
   $ E^+(\vec r;S)$ must 
scale as $ r^\zeta $ with $\zeta= (1-\alpha)/\nu +2-d-\eta$, namely
\begin{equation}
 E^+(\vec r;S) \approx E^+_0(S) r^{\zeta}
\label{E(r)} \end{equation}
for large enough $r$, independently of the spin and the lattice structure. 
 In 2D the  value 
 $\zeta=0.75$ is expected, while in 3D , adopting our recent 
estimate\cite{bcesse25} of the values of  
the correlation length exponent $\nu= 0.6299(2)$ and of the exponent
$\eta= 0.036(1)$, we should have $\zeta=0.3765(10)$.

\section{Numerical Results}



\begin{figure}
\includegraphics[width=4in]{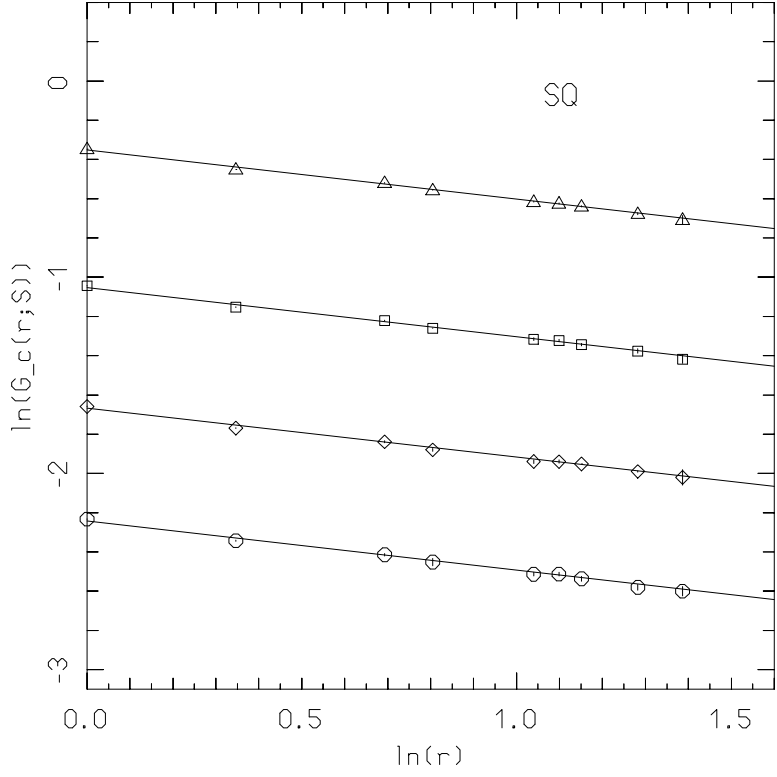}
\caption{\label{Gcsq}
Fig.1. Estimates of $G(\vec r,T_c;S)$ on the sq lattice.
The meaning of the symbols is as follows. Triangles: $S=1/2$, squares: $S=1$,
 rhombs: $S=3/2$, circles: $S=2$.
 The spin $S$ points are shifted vertically  by the quantity $1/2-S$
 in order to make the figure more legible.
 The continuous lines represent fits to the 
 leading asymptotic behaviors 
 ${\rm ln} G(\vec r,T_c;S) \approx c(S)-\eta {\ln} (r)$ 
 expected  for large $r$. We have taken $c(S)$ as a fit parameter 
 and fixed $\eta = 0.25$.}
\end{figure}

\begin{figure}
\includegraphics[width=4in]{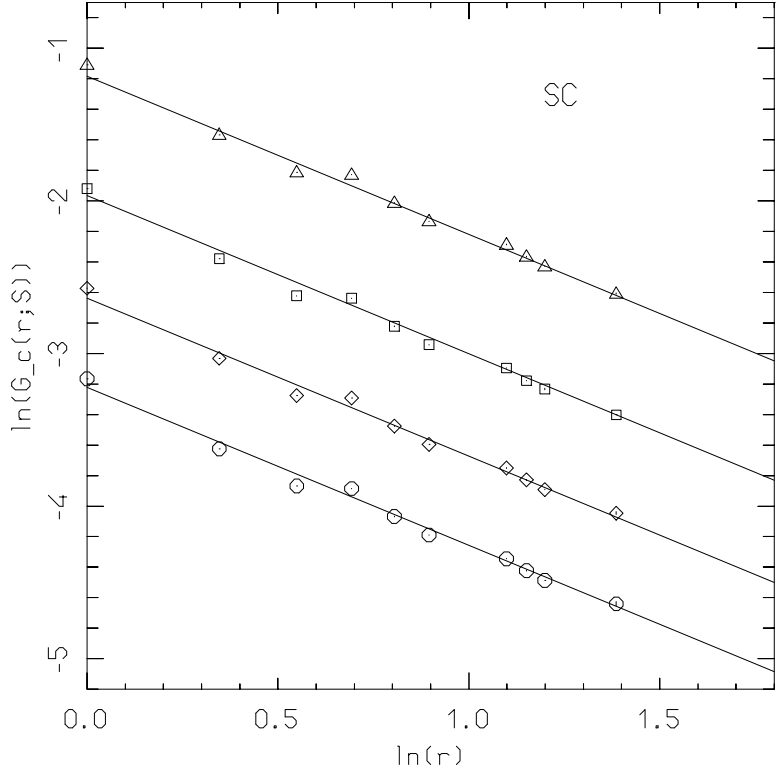}
\caption{\label{Gcsc}
Fig.2. Estimates of $G(\vec r,T_c;S)$ on the sc lattice.
The meaning of the symbols is the same as in Fig.\ref{Gcsq}
 The spin $S$ points are shifted vertically  by the quantity $1/2-S$  
  in order to make the figure more legible.
 The continuous lines represent fits to the 
 leading asymptotic behaviors  ${\rm ln} G(\vec r,T_c;S) \approx c(S)-
(1+\eta) {\ln} (r)$  expected  for large $r$.
We have taken $c(S)$ as a fit parameter and fixed  $\eta = 0.036$.}
\end{figure}

\begin{figure}
\includegraphics[width=4in]{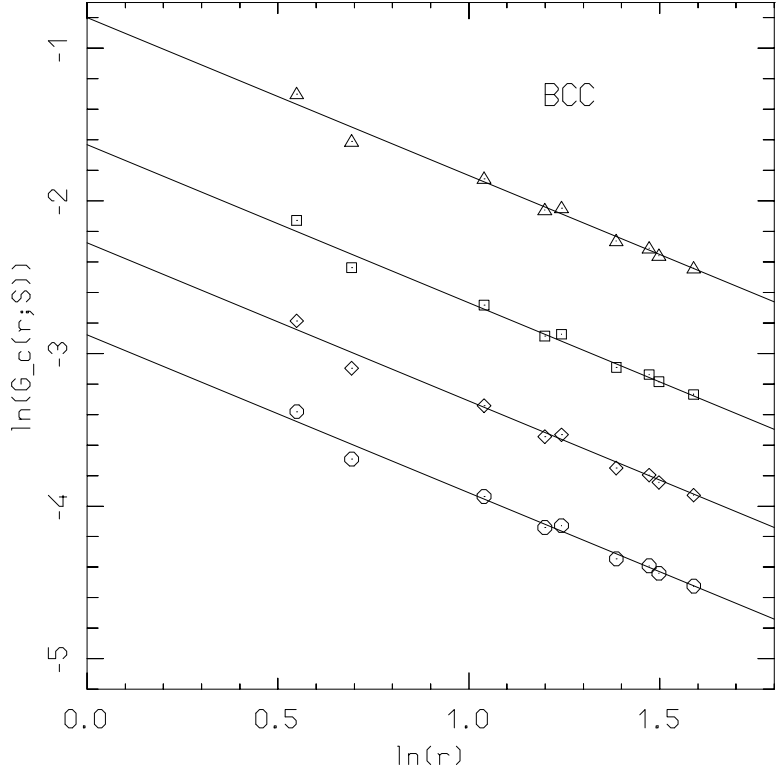}
\caption{\label{Gcbcc}
Fig.3. Estimates of $G(\vec r,T_c;S)$ on the bcc lattice.
The meaning of the symbols is the same as in Fig.\ref{Gcsq}.
 The spin $S$ points are shifted vertically  by the quantity $1/2-S$
 in order to make the figure more legible.
 The continuous lines represent fits to the  leading  
asymptotic behaviors  ${\rm ln} G(\vec r,T_c;S) \approx c(S)-
(1+\eta) {\ln} (r)$  expected  for large $r$.
We have taken $c(S)$ as a fit parameter and fixed  $\eta = 0.036$.}
\end{figure}

Let us first observe that, 
 due to the leading singular corrections 
in Eqs.(\ref{singpart2d}) and (\ref{singpart3d}), 
whose amplitudes $E^+(\vec r;S)$ grow with $r$  as  indicated by 
Eq.(\ref{E(r)}),
  determining accurately  $ G(\vec r,T_c;S)$ 
( as well as $E^+(\vec r;S)$ itself )
is a rather delicate matter for which it is crucial to rely on sufficiently 
many expansion coefficients.
We should  also consider that
 the number of non-trivial  coefficients in our series decreases 
with increasing $r$, and correspondingly the precision of our estimates of 
$G(\vec r,T_c;S)$ (and $E^+(\vec r,S)$) deteriorates. 
\begin{table}[!hbp]
\squeezetable
\caption{Our estimates of the critical-point values $ G(\vec r,T_c;S)$
of the spin-spin 
correlation function for the nearest-neighbor Ising models with  
spin $S=1/2,1,3/2,2,\infty$ on the sq, sc and bcc lattices.
For comparison with our results, the first column labelled $[S=1/2]$  
shows the only available previous estimates.
In the case of the sq lattice, we have cited the  exact\cite{fishbur,perk}
 values.
In the case of the nearest-neighbor 
 correlation ($r=(1,0,0)$) on the sc lattice, we have reported in the
first column the estimate obtained  
using our numerical procedure with 
the series $O(K^{45})$ of Ref.\cite{arisue} and have also cited
 a value\cite{luj} obtained in a recent high-precision MonteCarlo simulation. 
In the remaining
  cases we have reported the estimates of Ref.\cite{ritchie} 
obtained from series $O(K^{12})$.  We are not aware of previous
 calculations for $S>1/2$.} 
\label{tab1}
\begin{ruledtabular}
\begin{tabular}{lccccccc}
Lattice   &  $\vec r $  & [S=1/2] & S=1/2 & S=1& S=3/2& S=2& S=$\infty$ \\
\tableline  
sq& (1,0)&$0.707107..^a$&0.7071(1)& 0.5806(3)&0.517(1) &0.481(1)&0.338(1)\\
 &(1,1)&$0.636620..^a$&0.6366(1)&0.5207(4)&0.463(1)&0.431(1)&0.303(1)\\
 &(2,0)&$0.594715..^a$&0.5947(2)&0.486(1)& 0.433(1)&0.402(1)&0.282(2) \\
 & (2,1)&$0.573159..^a$ &0.573(1)&0.467(1)&0.417(2)&0.387(2)&0.272(2) \\
 & (2,2)&$0.540380..^a$&0.540(1)&0.442(1)&0.393(2)&0.365(2)&0.256(4)\\
sc &(1,0,0)& $0.330200(5)^b$ &0.33020(6)&0.24203(6)&0.20756(6)&0.18918(6)
&0.12886(6)\\
 &(1,0,0)&$0.33017(3)^c$&&&&&\\
 &(1,1,0)&$0.208(2)^d$ & 0.2086(1)&0.1529(1) &0.1311(1)&0.1194(1)&0.08141(5)\\
 &(1,1,1) &$0.164(4)^d$& 0.1633(1)&0.1197(1)&0.1027(1)&0.0936(1)&0.0638(1)\\
 &(2,0,0) &$0.162(4)^d$& 0.1608(2) &0.1178(2) &0.1010(2)&0.0921(2)&0.0627(1)\\
 &(3,0,0) &$0.104(7)^d$&0.1017(3)&0.0746(2)&0.0639(2)& 0.0581(2)&0.0396(1)\\
bcc&(1,1,1)&$0.2735(7)^d$& 0.27265(5)&0.19653(5)&0.16763(5)
&0.15243(5)&0.10341(5)\\
&(2,0,0)&$0.200(2)^d$&0.19971(5)& 0.14394(5)&0.12278(5)&0.11165(5)&0.07575(5)\\
&(2,2,0)&$0.157(2)^d$&0.15627(5)&0.11269(5)&0.09614(5)&0.08743(5)&0.05934(5)\\ 
&(3,1,1)&$0.129(3)^d$&0.12751(5)&0.09193(5)&0.07843(5)& 0.07132(5)&0.04839(5)\\
&(2,2,2)&$0.131(3)^d$&0.12914(5)&0.09315(5)&0.07945(5)&0.07224(5)&0.04903(5)\\
        
\end{tabular}
\end{ruledtabular}

$^a$Reference \cite{fishbur} $\quad$
$^b$Reference \cite{arisue}$\quad$ 
$^c$Reference \cite{luj}$\quad$
$^d$Reference \cite{ritchie} $\quad$

\end{table}
In Refs.\cite{fmw,fer,ritchie}, due to the small number  of coefficients 
available at that time,
a  generalized Neville extrapolation of the partial sums 
had to be used  
for determining $ G(\vec r,T;S)$ and $E^+(\vec r;S)$
in the vicinity of $T_c$.
Taking advantage of our new series, we can now 
  improve  substantially the numerical resummation 
of the HT series  by  resorting to first- or second-order  
inhomogeneous differential  approximants\cite{guttda}(DA's)   
biased with $K_c(S)$.
(Here and in what follows we have adopted the values of the critical 
temperatures 
 tabulated in Refs.\cite{bcesse25,bcesse2d25}.) 
 It does not come as a surprise that our procedures are  slightly 
less efficient in 2D than in 3D,  
 probably due to the presence of 
logarithms in the leading correction terms to the 
critical asymptotic behavior Eq.(\ref{singpart2d}) and  also that  in 3D
 the bcc lattice series always yield the most accurate results.
 If we restrict to  $ 1.< r < 6.$ 
the relative uncertainty of our  estimates of the critical sscf 
should generally remain well below $1\%$.
In 2D this can be guessed by comparing the estimates  of 
$G(\vec r,T_c;S)$ obtained  from our series $O(K^{25})$
 with the known exact results in the  sq lattice 
 case\cite{fishbur,perk} for $S=1/2$ and safely assuming that the  
precision does not deteriorate too fastly when higher values of $S$ 
are considered.
 In 3D no exact results are available, but 
the HT series for the nearest-neighbor correlation function
 was recently extended\cite{arisue}
 through order 45 in the  sc lattice case for $S=1/2$. 
 Therefore, in this case, we are  able to compare 
 our estimate  at order 25 
with the  result obtained by applying the same 
numerical procedures to the series 
 extended through order 45.
(It would be very interesting if the improved finite-lattice technique
 devised for this remarkable calculation 
 could be generalized as
effectively beyond first-neighbor correlations and  to general $S$.)
We should also mention that  a completely consistent alternative estimate
 of the critical sc-lattice nearest-neighbor sscf 
has been obtained\cite{luj}  in a recent high-precision MonteCarlo study.  
 For other values of $\vec r$ in the sc lattice case and
 in the bcc lattice case our results can only be  
 compared with  calculations\cite{ritchie}  using the old series $O(K^{12})$.
 Table I lists our estimates of $G(\vec r,T_c;S)$ with their apparent
 uncertainties  for a small sample of  values of $\vec r$ and $S$. 
Previous estimates of the critical sscf from shorter series, 
which  are available only for $S=1/2$, are shown  for comparison 
in the first column, labelled $[S=1/2]$, of this table. 
In Figs.\ref{Gcsq},\ref{Gcsc},\ref{Gcbcc} we have plotted  our estimates of
${\rm ln}\Big (G(\vec r,T_c;S)\Big )$  vs ${\rm ln} (r)$ 
for $ 1 \leq r \leq 5$ with $S=1/2,1,3/2,2$   
in the cases of the sq, sc and bcc lattices respectively.
 We have also shown by continuous lines
 the results of  one-parameter fits to the  leading asymptotic behaviors 
${\rm ln}\Big (G(\vec r,T_c;S)\Big ) \approx c(S)-(d-2+\eta){\rm ln} (r)$ 
expected for large $r$. We have taken only 
 $c(S)$ as a free parameter and fixed 
 $\eta=0.25$ in 2D and $\eta=0.036$ in 3D.  
 
Both in 2D and in 3D, we have
 estimated also $E^+(\vec r;S)$ from the amplitude of the singularity of the
 second temperature derivative of $G( \vec r,T;S)$, again using 
inhomogeneous first- and
second-order DA's biased with $ K_c(S)$ and $\alpha$. 
Our estimates of $E^+(\vec r;S)$ for a small sample of values of 
$\vec r$ and $S$ are shown in Table II. They are compared with the exactly
known values\cite{fishbur} for $S=1/2$, in the case of the sq 
lattice, or with a few
old estimates\cite{fishtar} from shorter series, 
in the case of the sc and bcc lattices.
 A comparison with  the exact results in 2D and with our estimate 
 using the mentioned  high-order
 calculation in the sc lattice\cite{arisue} for  $S=1/2$, 
 still suggests  that, for all values of $S$,
the relative accuracy of our  estimates should not be  generally 
 worse than $1\%$.

\begin{table}[!hbp]
\squeezetable
\caption{Amplitudes $E^+(\vec r;S)$ of the leading singular correction
 of the  sscf near the critical point 
for the nearest-neighbor Ising models with  
spin $S=1/2,1,3/2,2,\infty$ on the sq, sc and bcc lattices.
For comparison with our results,
 the first column of the table labelled $[S=1/2]$, shows  the available 
estimates from other sources.
In the case of the sq lattice, the exact values are taken 
from Ref.\cite{fishbur}.
In the case of the nearest-neighbor correlation on the sc lattice 
($r=(1,0,0)$),
we have reported in the first column our estimate obtained 
from the series $O(K^{45})$ of Ref.\cite{arisue}. In the remaining
  cases, whenever available, we have quoted  
 the estimates of Ref.\cite{fishtar} 
obtained from series $O(K^{12})$. We are not aware of  other published  
calculations for $S>1/2$.}
\label{tab2}
\begin{ruledtabular}
\begin{tabular}{lccccccc}
Lattice   &$\vec r $  & [S=1/2] & S=1/2 & S=1& S=3/2& S=2& S=$\infty$ \\
\tableline  
sq&(1,0)&$0.561100..^a$&0.562(1)&0.621(1)&0.623(1)& 0.613(1)&0.484(1)\\
 &(1,1)&$0.793515..^a$&0.794(1)&0.819(1)&0.812(1)&0.794(1)&0.616(1)\\
 &(2,0)&$1.0103348..^a$&1.01(1)&1.02(1)&1.01(1)&0.987(2)&0.759(2) \\
 &(2,1)&$1.120022..^a$&1.11(1)&1.13(1)&1.11(1)&1.08(1)&0.826(2) \\
sc&(1,0,0)&$2.252(5)^b$& 2.27(2)&2.16(2)&2.03(2)&1.93(2)&1.42(2)\\
 &(1,1,0)&$2.38(2)^c$& 3.01(2)&2.72(2)&2.52(2)&2.38(2)&1.72(2)\\
 &(1,1,1) &$2.86(4)^c$& 3.40(2)&3.03(2)&2.78(2)&2.62(2)&1.90(2)\\
 &(2,0,0) &$3.16(6)^c$& 3.53(2)&3.14(2)&2.88(2)&2.71(2)&1.95(2)\\
 &(3,0,0) &&4.36(2)& 3.79(2)&3.45(2)&3.24(2)&2.33(2)\\
bcc&(1,1,1)&$2.010^c$&2.325(5)&2.167(5)&2.022(5)&1.917(5)&1.401(5)\\
&(2,0,0)&& 2.707(6)& 2.442(6)& 2.256(6)&2.129(6)&1.545(6)\\
&(2,2,0)&&3.126(6)& 2.767(6)&2.535(6)&2.384(6)&1.720(6)\\ 
&(3,1,1)&&3.41(1)&2.98(1)&2.72(1)&2.55(1)&1.83(1)\\
&(2,2,2)&&3.44(1)&3.01(1)&2.74(1)&2.57(1)&1.84(1)\\        
\end{tabular}
\end{ruledtabular}
$^a$Reference \cite{fishbur} $\quad$
$^b$Reference \cite{arisue}$\quad$
$^c$Reference \cite{fishtar}
\end{table}

\begin{figure}
\includegraphics[width=4in]{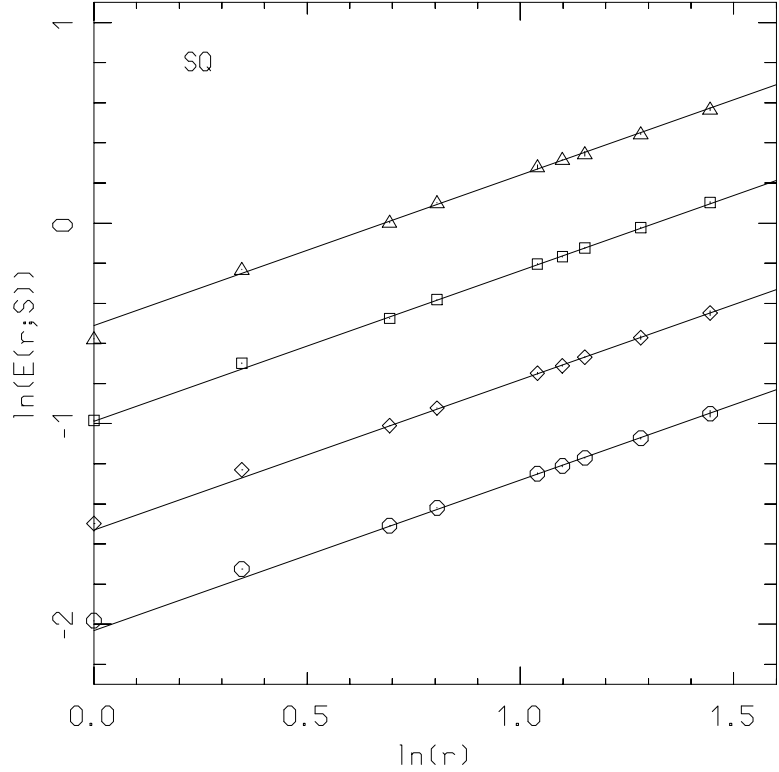}
\caption{\label{fig_E_sq} 
Fig.4. Estimates of $E^+(\vec r;S)$ on the sq lattice.
The meaning of the symbols is the same as in Fig.\ref{Gcsq}.
 The spin $S$ points are shifted vertically  by the quantity $1/2-S$
 in order to make the figure more legible.
 The continuous lines represent fits to the  leading  
asymptotic behaviors ${\rm ln} E^+(\vec r;S) \approx b(S) +{\zeta}{\rm ln}(r)$
 expected  for large $r$. We have taken $b(S)$ as a fit parameter 
and fixed  $\zeta = 0.75$.}
\end{figure}

\begin{figure}
\includegraphics[width=4in]{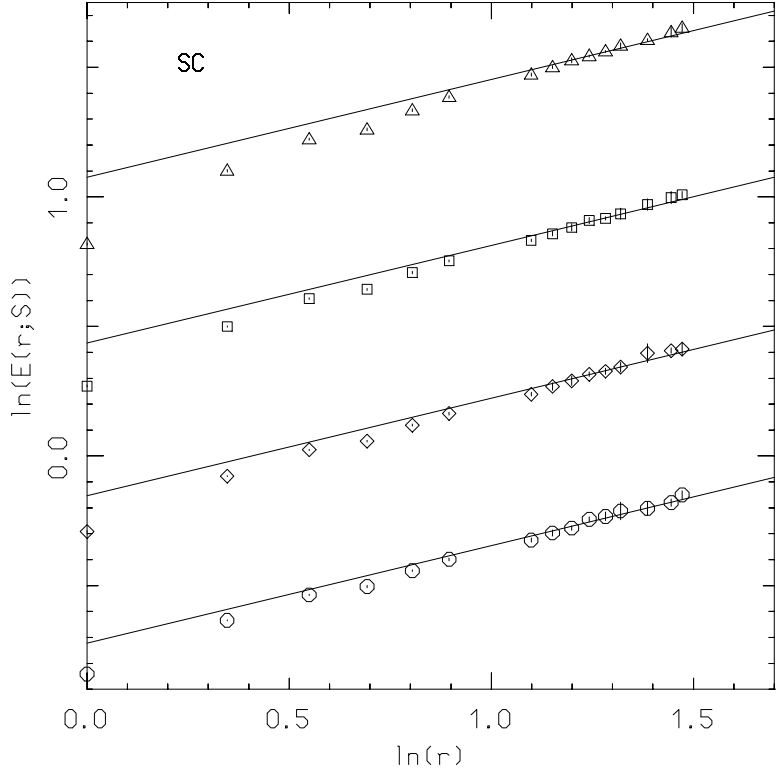}
\caption{\label{Esc} 
Fig.5. Estimates of $E^+(\vec r;S)$ on the sc lattice.
The meaning of the symbols is the same as in Fig.\ref{Gcsq}.
 The spin $S$ points are shifted vertically  by the quantity $1/2-S$
 in order to make the figure more legible. 
The continuous lines represent fits to the  leading 
asymptotic behaviors ${\rm ln} E^+(\vec r;S) \approx b(S) +{\zeta}{\rm ln}(r)$
  expected  for large $r$. We have taken
 $b(S)$ as a fit parameter and fixed $\zeta = 0.3765$.}
\end{figure}

\begin{figure}
\includegraphics[width=4in]{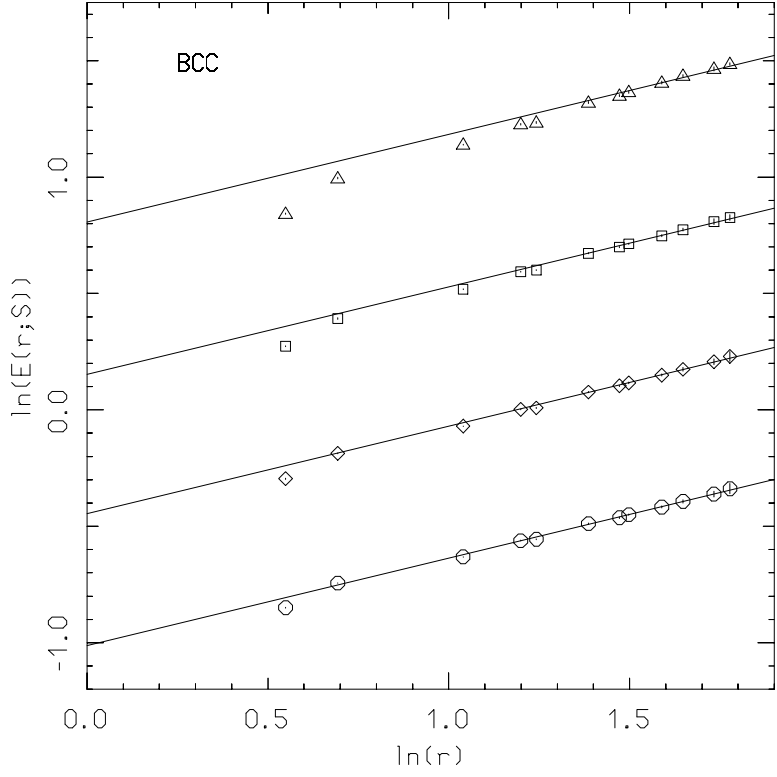}
\caption{\label{Ebcc} 
Fig.6. Estimates of $E^+(\vec r;S)$ on the bcc lattice.
The meaning of the symbols is the same as in Fig.\ref{Gcsq}.
 The spin $S$ points are shifted  vertically by the quantity $1/2-S$
 in order to make the figure more legible. 
The continuous lines represent fits to the  leading 
asymptotic behaviors ${\rm ln} E^+(\vec r;S) \approx b(S) +{\zeta}{\rm ln}(r)$
 expected  for large $r$.
We have taken $b(S)$ as a fit parameter and fixed  $\zeta = 0.3765$.} 
\end{figure}

In Figs. \ref{fig_E_sq},\ref{Esc},\ref{Ebcc} for $ S=1/2,1,3/2,2$ 
 we have plotted,  
${\rm ln}\Big (E^+(\vec r;S)\Big )$ vs ${\rm ln}(r)$ 
  in the case of the sq, sc, and bcc lattices respectively. 
For $r > 4.5 $ in the case of the sc lattice and $r > 6. $ in the case
of the bcc lattice, we have not reported any estimates of $E^+(\vec r;S)$, 
 because the available nontrivial HT coefficients of the sscf are not 
sufficiently many to allow  estimates at the level of precision above
mentioned. 
In these figures we have also represented 
by continuous lines the results of  one-parameter fits to 
the leading asymptotic behaviors 
${\rm ln}\Big (E^+(\vec r;S)\Big ) \approx b(S) + \zeta {\rm ln}(r)$  
expected for large $r$.  
We have taken  for $\zeta$  the 
expected values $\zeta= 0.75$ in the 2D case and  $\zeta= 0.3765$ in the 
3D cases, while the free parameters $b(S)$ have been determined   
 using in the fits only the data with $r \gtrsim 1.8$. 
 Indeed, our new data show   visible deviations from asymptotic scaling  
  for sufficiently small $r$,  particularly so in the case of
 the sc lattice, but the asymptotic consistency with the strong-scaling
 hypothesis Eq.(\ref{E(r)}) is good. 
 The behavior of $E^+(\vec r;1/2)$
 as a function of $r$ was first studied in Ref.\cite{fmw}
 using series $O(K^{12})$ for the face-centered cubic lattice. 
 In that analysis both $\zeta$ and  $b(1/2)$ were 
determined by a two-parameter fit of the numerical results 
to the  leading asymptotic behavior 
 under the very 
simple assumption that the corrections to scaling are negligible
 even for  `not very large' $r$. 
(As we have indicated above,  our new data show that such a 
strong assumption is untenable.)
The authors of Ref.\cite{fmw} concluded  
 that $\zeta = 0.47(6)$, an estimate in sharp disagreement with the value
$\zeta = 0.33(1)$ expected from the exponent values $\nu= 0.638(2)$ and 
$\eta = 0.041(6)$ generally accepted at that time.
 A few years later,  for $S=1/2$,  the somewhat lower
 estimate $\zeta = 0.39(4)$ was obtained\cite{fishtar}
 from an analysis of  LT expansions in powers of 
$u=exp(-2K)$ up to order $u^{11}$ and $u^{13}$, on the sc and the 
bcc lattices respectively. In this latter study, however,
  a third fit parameter 
 had been  introduced in order 
 to allow for small corrections to the  asymptotic scaling 
behavior of $E^+(\vec r;S)$.
Also this estimate of $\zeta $
did not agree with the value expected at that time, 
but is  quite compatible with the  presently preferred value.  

 Before  any strong confidence  
in the results of such two- or three-parameter fits can be 
justified, we believe, however, that 
the HT  series should be further extended  in order  
 to enlarge significantly the range  of values of $r$ for which  
$E^+(\vec r;S)$ can be determined with sufficient accuracy.

\begin{figure}
\includegraphics[width=4in]{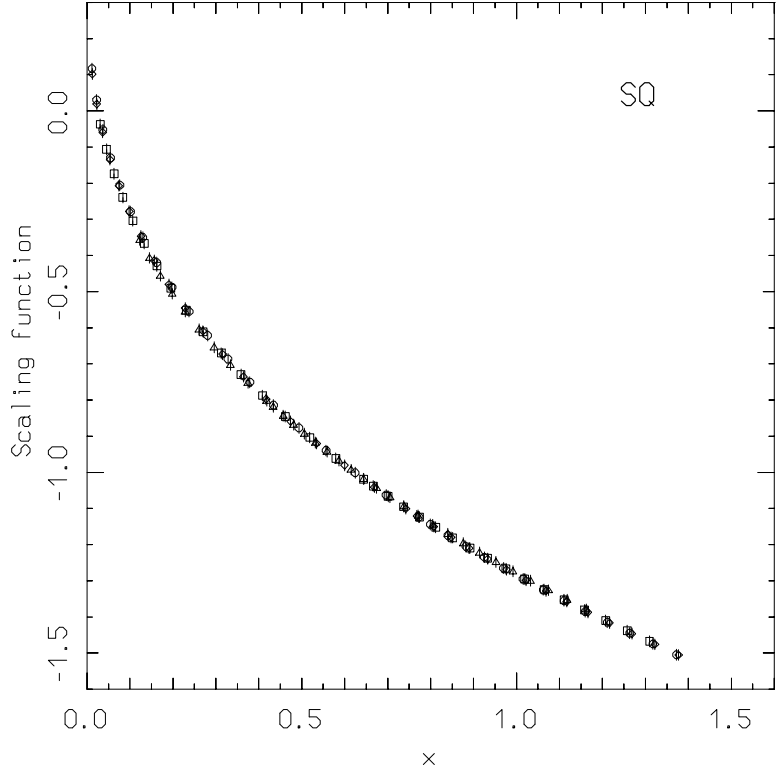}
\caption{\label{scalingsq} 
Fig.7. The logarithm of the scaling function 
$r^{d-2+\eta}G( \vec r,T;S)$ vs. $x=r/\xi_{sm}(T;S)$ 
in the case of the sq lattice. The data represent the
ssfc's with $\vec r=(2,0)$ and $S=1/2,1,3/2,2$ in the range of temperatures
 for which $1.5 \lesssim \xi(T;S)\lesssim  200$.
The meaning of the symbols is the same as in Fig.\ref{Gcsq}.}
\end{figure}

\begin{figure}
\includegraphics[width=4in]{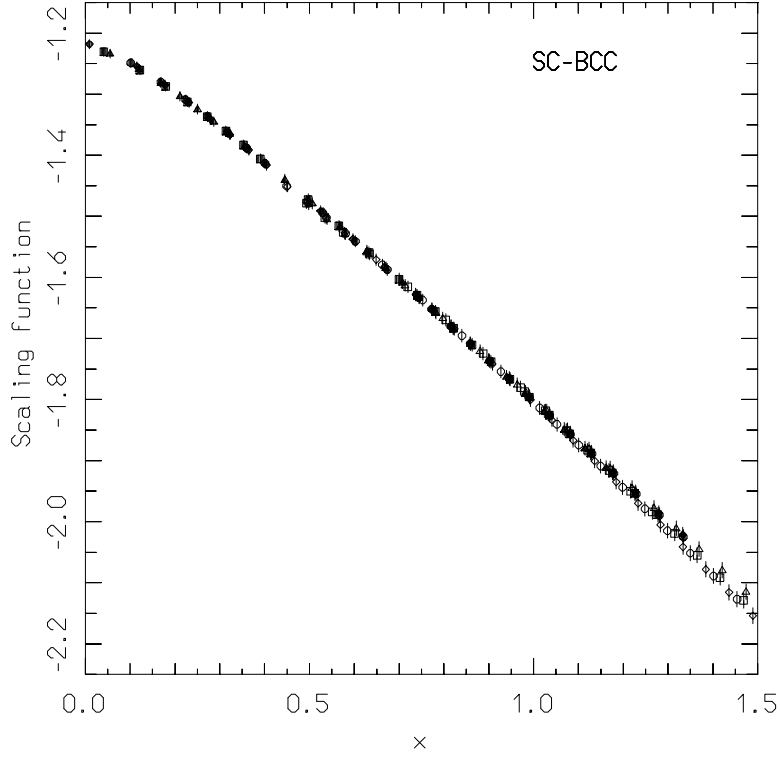}
\caption{\label{scalingscbcc} 
Fig.8. The logarithm of the scaling function 
$r^{d-2+\eta}G( \vec r,T;S)$ vs. $x=r/\xi_{sm}(T;S)$ 
in the case of the sc and the bcc lattices. For both lattices 
 the data represent the 
ssfc's with $\vec r=(4,0,0)$ and $S=1/2,1,3/2,2$ 
in the range of temperatures for which $2.7 \lesssim \xi(T;S) \lesssim  400$.
The meaning of the symbols is the same as in Fig.\ref{Gcsq} for the sc
lattice case. For the bcc lattice data we have used full symbols of
the same shape.}
\end{figure}

Having tabulated a wide sample of estimates of $G(\vec r,T_c;S)$ and
 $E^+(\vec r;S)$ 
 with some improvement both in the extent and 
 the accuracy,  with respect to the very few estimates available in the
 literature,
we are now in the position to
 exhibit more directly the scaling property by  
examining the near-critical sscf in the $r$-space.
 For  $T \rightarrow T_c+0 $,  as suggested by Eq.(\ref{scalfun}),  
by a proper choice of the 
scale factors $A_l(S)$ and
$ C_l(S)$,
 we should be able to plot the quantities 
\begin{equation}
r^{d-2+\eta}G( \vec r,T;S) \approx A_l(S) {\rm D_0}\Big 
(C_l(S) r/\xi_{sm}(T;S)\Big ) 
\label{funfig}\end{equation}
vs. $r/\xi_{sm}$ in such a way that the  curves, associated to  various 
 values of $S$ and to different lattices, collapse on each other.
In Fig.\ref{scalingsq},
we have plotted 
 ${\rm ln} \Big ( r^{d-2+\eta}G( \vec r,T;S)\Big )$ vs. $r/\xi_{sm}(T;S)$ 
in the case of the sq lattice
taking $\vec r = (2,0)$ and  $S=1/2,1,3/2,2$.
 Our data points refer to the range of temperatures
for which $1.5 \lesssim \xi_{sm} \lesssim 200$. 
 Fig. \ref{scalingscbcc} shows the analogous plot for 
${\rm ln}\Big ( r^{d-2+\eta}G( \vec r,T;S)\Big )$ vs. $r/\xi_{sm}$ 
in the case of the sc and bcc lattices. Here we have taken $\vec r = (4,0,0)$ 
 and $S=1/2,1,3/2,2$ and have 
plotted data in the range of  temperatures
for which $2.7 \lesssim \xi_{sm} \lesssim  400 $. 
Completely consistent results are obtained also for other choices of
 $\vec r$.  
 As already observed, within  these limitations, the present length 
of the HT series  appears sufficient to 
obtain reliable estimates 
and our results are consistent with
 the  strong-scaling hypothesis to a good approximation.

The very small mismatch of the curves in the extreme regions
$r/\xi_{sm} \ll 1$ or $r/\xi_{sm} \gg 1$ which can still be observed 
 is related:  
i) to the fact that the scaling 
property has an asymptotic character, while in 
practice the size of $r$ still cannot  
exceed a few lattice spacings if we want 
to use a  decent
number of expansion coefficients  in the estimate
of $G( \vec r,T;S)$, 
ii) to the residual
influence of the  subcritical corrections.

\begin{table}[!hbp]
\squeezetable
\caption{Estimates of $\Sigma_4(T_c;S)$ (see Eq.(\ref{sigma4}))
 and $\Sigma_6(T_c;S)$ (see Eq.(\ref{sigma6})) 
 in the case of the sq, sc and bcc lattices for various values
of $S$. For comparison, we have also reported  a few previous estimates   
 listed in Ref.\cite{martin,pisa}, indicating the method of calculation 
and the expected uncertainty, when available. Only for convenience, 
we have listed in the $S=1/2$ column  also the results of the 
renormalization group  and the optimized continuous spin calculations 
which, of course, do not refer to spin $S=1/2$.}
\label{tab3}
\begin{ruledtabular}
\begin{tabular}{lcccccc}
 Quantity & Lattice & S=1/2 & S=1& S=3/2& S=2& S=$\infty$ \\
\tableline  
$\Sigma_4(T_c;S)\times 10^{4}$&sq&7.8(3)& 7.9(3)& 7.9(3)&7.6(3)&7.5(3) \\
$\Sigma_4(T_c;S)\times 10^{4}$(Exact)$^a$&sq&7.936796\ldots &&&&\\
$\Sigma_6(T_c;S)\times 10^{5}$&sq&1.1(1)&1.1(1)&1.0(1)&1.1(1)&1.0(1) \\
$\Sigma_6(T_c;S)\times 10^{5}$(Exact)$^a$&sq&1.095991\ldots&&&&\\
\tableline
$\Sigma_4(T_c;S)\times 10^{4}$&sc&3.76(8)&3.9(2)&3.77(8)&3.75(8)
&3.7(2)\\
$\Sigma_4(T_c;S)\times 10^{4}$&bcc&3.75(5)&3.74(5)&3.76(5)&3.76(5)
&3.77(5)\\
$\Sigma_6(T_c;S)\times 10^{5}$&sc &1.0(2)&.9(2)&0.9(2)&0.8(2)&0.7(2)\\
$\Sigma_6(T_c;S)\times 10^{5}$&bcc&0.9(1)&0.86(5)&0.85(5)&0.85(5)&.85(5)\\
$\Sigma_4\times 10^{4}$[HT]$^b$&sc &3.0(2) &&&&\\
$\Sigma_4\times 10^{4}$[HT]$^c$&sc &5.5(15) &&&&\\
$\Sigma_4\times 10^{4}$[HT]$^c$&bcc &7.1(15) &&&&\\  
$\Sigma_6\times 10^{5}$[HT]$^b$&sc &0.5(2)&&&&\\
$\Sigma_6\times 10^{5}$[HT]$^c$&sc &0.5(2)&&&&\\
$\Sigma_6\times 10^{5}$[HT]$^c$&bcc &0.9(3)&&&&\\
$\Sigma_4\times 10^{4}$ [opt.cont.spin]$^b$&sc& 3.90(6) & & & &\\
$\Sigma_6\times 10^{5}$ [opt.cont.spin]$^b$&sc& .88(1) & & & &\\
$\Sigma_4\times 10^{4}$ [$\epsilon$-expans.]$^b$&&3.3(2) & & & &\\
$\Sigma_6\times 10^{5}$ [$\epsilon$-expans.]$^b$&&0.7& &  & &\\
$\Sigma_4\times 10^{4}$ [g-expans.]$^b$&&4.0(5)& & & &\\
$\Sigma_6\times 10^{5}$ [g-expans.]$^b$&&1.3(3)&  & &\\
\end{tabular}
\end{ruledtabular}
 $^a$ Reference \cite{pisa}$\quad$
$^b$Reference \cite{martin} $\quad$ $^c$Reference \cite{fishbur}  
\end{table}

We can further test the universality properties of the 
sscf in the $k$-space, namely  the critical 
scattering function, by simply 
showing that $\Sigma_4(T_c;S)$ and $\Sigma_6(T_c;S)$
 are independent of $S$ and of the lattice structure.
 Also these quantities are calculated by first- and second-order
DA's biased with $K_c(S)$. Since higher-order moments of the sscf
 (in which the less accurately known correlations between distant
 spins are weighted much more than those between near spins) 
enter into the definitions eq.(\ref{sigma4}) and eq.(\ref{sigma6}), 
the convergence of the extrapolations is not expected to be 
very fast, particularly so
 in the cases of the sq and sc lattices. 
 We should also 
 consider that $\Sigma_4(T_c;S)$ is
the very small difference   between unity and the critical value of
 some multiple of  a ratio of moments of the sscf, 
so that a very high accuracy 
in the estimate of the latter is 
needed to achieve even a relatively modest precision 
for $\Sigma_4(T_c;S)$. The same remark applies 
also  in the case of $\Sigma_6(T_c;S)$.
In Table \ref{tab3} we have collected our estimates of  
$\Sigma_4(T_c;S)$ and $\Sigma_6(T_c;S)$  in the case of the sc, sq and 
bcc lattices for $S=1/2,1,3/2,2,\infty$. We have also reported  
 a few previous estimates\cite{martin,pisa} from the existing literature.

In the case of the sq lattice our data suggest   
the final estimates $\Sigma_4(T_c;S)=7.8(3)\times 10^{-4}$ 
and $\Sigma_6(T_c;S)=1.1(1)\times 10^{-5} $, 
independently of $S$ and in reasonable agreement with the high-precision
determinations\cite{pisa} 
of $\Sigma_4(T_c;1/2)=7.936796\ldots \times 10^{-4} $ 
and of $\Sigma_6(T_c;1/2)= 1.095991\dots \times 10^{-5}$ 
obtained by numerical integration of the analytically known\cite{wu} sscf
of the  $S=1/2$ model in 2D.
In 3D our results for the bcc lattice 
show a definitely smaller uncertainty than for the sc lattice. 
They  suggest the 
final estimates  $\Sigma_4(T_c;S)=3.8(1)\times 10^{-4}$ 
  and $\Sigma_6(T_c;S)=0.9(1)\times 10^{-5} $ 
independently of the spin $S$ and lattice structure. Our results are 
therefore consistent with
the corresponding estimates in the literature, 
 in particular with the values $\Sigma_4(T_c)= 3.90(6)\times 10^{-4} $ 
and $\Sigma_6(T_c)= 0.88(1) \times 10^{-5}$  obtained\cite{martin,pisa} 
optimizing the parameters of a continuous-spin model, under the assumption
 of universality.
Let us also mention that renormalization group 
 calculations\cite{martin} in the  
$\epsilon$-expansion  scheme to third order yielded the 
 estimates $\Sigma_4=3.3(2)\times 10^{-4} $ 
and $\Sigma_6= 0.7 \times 10^{-5} $, while, in the 
 coupling-constant expansion scheme to fourth order, 
the corresponding results 
were $\Sigma_4=4.0(5)\times 10^{-4} $ 
and $\Sigma_6= 1.3(3) \times 10^{-5} $.

\begin{table}[!hbp]
\squeezetable
\caption{Estimates of the moment ratios $R_{m,n;r,s}(T_c;S)$ 
(see Eq.(\ref{Rmnrs})) 
 in the case of the sq, sc and bcc lattices for various values
of $S$.}
\label{tab4}
\begin{ruledtabular}
\begin{tabular}{lccccccc}
$R_{m,n;r,s}$&Lattice   & S=1/2 & S=1& S=3/2& S=2& S=$\infty$ \\
\tableline  
$R_{0,1;1/2,1/2}$&sq &1.1641(1)&1.1642(1)&1.1642(1)&1.1642(1)&1.1641(1)\\
$R_{0,1;1/4,3/4}$&sq &1.1211(1)&1.1211(1)&1.1211(1)&1.1211(1)&1.1210(1)\\
$R_{-3/4,1/4;-1/4,-1/4}$&sq &1.299(1)&1.300(1)&1.301(1)&1.300(1)
&1.301(1)\\
$R_{-1,-1/2;-3/4,-3/4}$&sq &1.121(5)&1.124(4)&1.124(4)&1.125(4)&1.126(4)\\
\tableline 
$R_{0,1;1/2,1/2}$ &sc &1.1320(1)&1.1320(2)&1.1319(1)
&1.1319(2)&1.1319(2)\\
$R_{0,1;1/2,1/2}$&bcc &1.1320(1)&1.1319(1)&1.1319(1)&1.1319(1)&1.1319(1)\\
$R_{0,1;1/4,3/4}$ &sc &1.0977(2)&1.0977(2)&1.0976(2)
&1.0976(2)&1.0976(2)\\
$R_{0,1;1/4,3/4}$&bcc &1.0977(1)&1.0976(1)&1.0976(1)&1.0976(1)&1.0976(1)\\
$R_{1/2,1/2;1/4,3/4}$&sc &0.9697(2)&0.9697(2)&0.9698(2)&0.9697(2)
&0.9697(2)\\
$R_{1/2,1/2;1/4,3/4}$&bcc &0.9697(1)&0.9697(1)&0.9697(1)&0.9697(1)
&0.9698(1)\\
$R_{-1,-1/2;-3/4,-3/4}$&sc &1.084(1)&1.084(1)&1.084(1)&1.084(1)&1.083(1)\\
$R_{-1,-1/2;-3/4,-3/4}$&bcc &1.083(1)&1.083(1)&1.083(1)&1.083(1)&1.083(1)\\
\end{tabular}
\end{ruledtabular}
\end{table}

The results of our analysis of the universal ratios $R_{m,n;r,s}(T_c;S)$ 
 are reported in table \ref{tab4}. They also show independence 
of the spin and   of the lattice structure within a good precision. 
 Our series-extrapolation procedure, based on first- and 
second-order DA's uses only our estimates of $K_c(S)$ and does 
not need to be biased also with $\gamma$ and $\nu$ 
as it was necessary in the generalized Neville 
procedure\cite{fmw,fer} 
 employed with the short series of Ref.\cite{fermor}.
Considering  that the values $\gamma=1.25$ and $\nu=0.625$ (or 
$\nu=0.638$) of the exponents accepted at the time of that study are somewhat 
different from the currently preferred ones and that the extrapolations
 are very sensitive to those values, a comparison with the 
numerical results of Ref.\cite{fermor} has little meaning.

Finally, we have  tested both in 2D and in 3D 
 the   spin  independence of the ratio $Q^+_{\xi}(S)$
 defined by Eq.\ref{ratio}. In 3D  also  the 
lattice independence of $Q^+_{\xi}(S)$ can be tested.

In 2D, on the sq lattice, the non-trivial expansion 
coefficients of the ratio $\xi^2_{\vec e}(T;S) / \xi^{2}_{sm}(T;S)$
  are not sufficiently
 many and their behavior is not smooth enough 
to yield very accurate results. 
Therefore our best estimate of $Q^+_{\xi}(S)$
  (by first-order DA's  biased with $K_c(S)$) 
cannot be  more precise than $Q^+_{\xi}(S)=1.0004(2)$, 
independently of $S$. 
Our rough estimate is, however, consistent with   the more accurate
 determination  $Q^+_{\xi}(1/2)=1.000402 \ldots$  obtained in 
 the $S=1/2$ case in which, as already indicated above, 
very long series are available\cite{nick21}
 for $\xi^2_{sm}(T;1/2)$, while 
$ \xi^2_{\vec e} (T;1/2) $ is exactly known\cite{onsa,fishbur}. 

In 3D we can use both first- and second-order DA's biased with $K_c(S)$.
 The very smooth 
bcc lattice series yield the most accurate results. Our final estimate is   
$Q^+_{\xi}(S)= 1.000200(3)$, independently of $S$ and of the lattice structure.
 So far, 
 this ratio could  be computed\cite{martin} only for $S=1/2$
 from a 15 term series on the sc lattice, with the 
result 
$ Q^+_{\xi}(1/2)=1.000125(50)$. A  more precise estimate\cite{martin}  
$Q^+_{\xi}=1.000199(3)$ 
 was obtained indirectly (and assuming universality),
  from  optimized HT series  for a continuous-spin model 
on the sc lattice.     
Within the renormalization group approach\cite{martin},
 the estimate $Q^+_{\xi}=1.000160(20)$ 
was obtained in
  the $\epsilon$-expansion  to third order, 
while the coupling-constant expansion technique to fourth order gave   
$ Q^+_{\xi}= 1.000205(30) $.

\acknowledgments
The second named author (M.C.)  passed away
 before the final text of this report was completed, therefore 
the first author is entirely responsible for any errors or omissions.

This work has been partially supported by the Ministry of  
University and Research.

\newpage

\end{document}